\documentclass[prb,citeautoscript,twocolumn]{revtex4}
\usepackage{amssymb}
\usepackage{epsfig}
\usepackage{dcolumn}
\begin{document}

\title{Comment on ''First-principles calculation of the superconducting
transition in MgB$_{2}$ within the anisotropic Eliashberg formalism''}
\author{I. I. Mazin$^1$}
\author{O. K. Andersen$^2$}
\author{O. Jepsen$^2$}
\author{A. A. Golubov$^3$}
\author{O. V. Dolgov$^{2,4}$}
\author{J. Kortus$^2$}
\affiliation{$^{1}$Center for Computational Materials Science, Naval Research Laboratory,
Washington, DC 20375-5000, USA\\
$^2$Max-Planck-Institut f{\"u}r Festk{\"o}rperforschung, Heisenbergstr. 1,
D-70569 Stuttgart, Germany\\
$^{3}$University of Twente, Department of Applied Physics, 7500 AE Enschede,
The Netherlands\\
$^{4}$Eberhard-Karls-Universit\"{a}t, Auf der Morgenstelle 14, D-72076 T\"{u}%
bingen, Germany}
\date{\today}

\begin{abstract}
Choi \textit{et al.} [Phys. Rev. B 66, 020513 (2002)] recently presented first
principles calculations of the electron-phonon coupling and
superconductivity in MgB$_{2}$, emphasizing the importance of anisotropy and
anharmonicity. We point out that (1) variation of the superconducting gap
inside the $\sigma$- or the $\pi$-bands can hardly be observed in real samples,
and (2) taking the anisotropy of the Coulomb repulsion into account
influences the size of the small gap, $\Delta _{\pi }.$
\end{abstract}

\pacs{63.20.Kr,74.20.-z,74.70.Ad,78.20.Bh}
\maketitle

In a recent paper \cite{Louie}, as well as in a follow-up paper \cite{Choi},
Choi \textit{et al.} presented an \textit{ab initio} calculation of the superconducting
transition and superconducting properties of MgB$_{2}$. The important
improvement over existing calculations was that they allowed the order
parameter to vary freely over the Fermi surface, i.e. $\Delta =\Delta \left( 
\mathbf{k}\right)$, and at the same time took the anharmonicity into
account. As a consequence, they had to compute the fully anisotropic
electron-phonon interaction, $\lambda \left( \mathbf{k,k}^{\prime }\right)$,
and to solve the corresponding Eliashberg equation. The Coulomb
pseudopotential, $\mu^{\ast }\left( \mathbf{k,k}^{\prime }\right)$, was
assumed not to depend on $\mathbf{k}$ and $\mathbf{k}^{\prime }$,
and was treated as an adjustable parameter. First-generation \textit{ab initio}
calculations of the superconducting transition and superconducting
properties of MgB$_{2}$ had assumed $\Delta $ to be constant and had
therefore solved merely the isotropic Eliashberg equation \cite{Kong}.
Moreover, anharmonicity had been neglected. It was soon pointed out 
\cite{Amy} that
the calculated electron-phonon coupling suggests that the gap 
on the two $\pi $ sheets of the Fermi
surface is smaller than that on the two $\sigma $ sheets, and that
anharmonicity is important. This led to the so-called two-band model. 
Ab initio calculations of the second generation \cite%
{Amy,GolubovJPCM,Brinkman} allowed for two, and sometimes four  gaps, $%
\Delta _{n},$ and thus had to compute $\lambda _{nn^{\prime }}$ , to
estimate the anisotropy of $\mu _{nn^{\prime }}^{\ast },$ and to solve the
corresponding Eliashberg equations.

Here we shall comment on (1) whether consequences of anisotropy beyond that
of the two-band model may be observed and (2) whether at this level of
detail Choi \textit{et al.}'s assumption of a uniform Coulomb repulsion is warranted.

1. Ref. \onlinecite{Louie} implies that there is a distribution of gaps 
\emph{within} the $\sigma $ and the $\pi $-sheets, not only in the
calculations for perfectly clean MgB$_{2}$, but also in the actual material;
in other words, that the distribution of gaps shown in Fig. 2 of Ref. %
\onlinecite{Choi} is \emph{observable.}\textit{\ }However, in the theory of
anisotropic superconductivity it is known that any intraband nonuniformity
of the order parameter is suppressed by strong intraband impurity scattering.
It is not immediately obvious, though, when scattering should be considered 
strong in this connection.
Since excitation gaps are not equal to the order parameters any more, 
one needs to compare individual densities of states (DOS), $N(E)$, 
for the two $\sigma$-bands (or the two $\pi$-bands) 
with each other for a given scattering strength, and check 
whether $|N_{\sigma 1}-N_{\sigma 2}|\ll |N_{\sigma 1}+N_{\sigma 2}|$. 
The relevant expression can be found in Ref.~\onlinecite{Golubov}. 
In the limit of large scattering rates, $\gamma$, one can derive an analytical
expression for this criterion, namely $\gamma > \sqrt{\left\langle \Delta
\right\rangle \cdot \delta \Delta }$ (Ref.~\onlinecite{gaps}), where 
$\left\langle \Delta \right\rangle $ is the average \textit{order parameter},
and $\delta \Delta $ is the variation of the \textit{order parameter} over
the Fermi surface in question. 
With the data from Refs.~\onlinecite{Louie,Choi} for $\langle\Delta\rangle$ and $\delta\Delta$,
this gives characteristic scattering rates of respectively 2 and 1.5 meV for
the $\sigma $ and $\pi $-bands. 
Therefore, to observe 4 distinct gaps in MgB$_{2}$ one needs samples with 
scattering rates smaller than 2 meV, that is, with mean free paths beyond 1500 \AA. 
To observe gap variations beyond the
4-band model, far cleaner samples are needed. This is the reason why
at most two distinct gaps have been observed in experiments. It is
even surprising that the difference of 5 meV between the gaps of the $\sigma 
$ and the $\pi $-bands is not smeared out. This seems to be due to the 
inability of common impurities to couple between the disparate $\sigma $ and 
$\pi $-band wavefunctions\cite{Mazin}, so that $\gamma _{\sigma \pi }\ll
\gamma _{\sigma \sigma }\sim \gamma _{{\pi \pi }}$.

2. For the Coulomb pseudopotential, Choi \textit{et al.} used $\mu ^{\ast }(%
\mathbf{k,k}^{\prime })=\mu ^{\ast }(\omega _{c})=0.12$ (with the cut-off
frequency $\omega _{c}\approx 5\omega _{ph}^{\max })$ and stated that the
superconducting properties of MgB$_{2}$ were not very sensitive 
to the choice
of $\mu ^{\ast }(\omega _{c}).$ This at first seems plausible, because the 
Coulomb pseudopotential enters the Eliashberg equation only in the
combination $\lambda (\mathbf{k,k}^{\prime },\nu -\nu ^{\prime })-\mu ^{\ast
}(\mathbf{k,k}^{\prime }),$ and the $\lambda $-distribution varies on the
scale of $\sim $1.8, $\sim $0.3, and $\sim $0.2 for $\sigma \sigma $-, $\pi
\pi $-, and $\sigma \pi $-scattering  respectively [see Fig. 3 of Ref. %
\onlinecite{Louie}]. Therefore, at most the $\sigma \pi $-scattering can be
influenced by anisotropy of $\mu ^{\ast }$. We shall argue that the $\sigma
\pi $-interband Coulomb matrix elements \emph{are} considerably smaller than
the intraband matrix elements due to the very small overlap of the $\sigma $%
- and $\pi $-band charge densities\cite{GolubovJPCM} and that this is
sufficient to influence the superconducting properties, in particular the
size of the small gap, $\Delta _{\pi }$.

Choi \textit{et al.} do not give the band-integrated values of their
coupling constants, but by integrating Fig. 3 of Ref. \onlinecite{Louie}
with the DOS-ratio $N_{\pi }/N_{\sigma }=1.37$ according to:%
\begin{eqnarray}
\lambda _{nn^{\prime }}(0) &\equiv &\frac{1}{N}\sum_{\mathbf{k,k}^{\prime }}%
\frac{\delta (\varepsilon _{n\mathbf{k}})}{N_{n}}\lambda (\mathbf{k},\mathbf{%
k}^{\prime }\mathbf{,}0)\delta (\varepsilon _{n^{\prime }\mathbf{k}^{\prime
}})  \label{e1} \\
&=&\frac{1}{N}\sum_{\mathbf{k,k}^{\prime }}W_{n\mathbf{k}}\,\lambda (\mathbf{%
k},\mathbf{k}^{\prime }\mathbf{,}0)W_{n^{\prime }\mathbf{k}^{\prime
}}N_{n^{\prime }}  \nonumber
\end{eqnarray}%
for the phonon-mediated coupling of an electron in band $n$ to all electrons
in band $n^{\prime },$ we can map Choi \emph{et al.}'s fully anisotropic
model onto a two-gap model with $\lambda _{\sigma \sigma }=0.78,\;\lambda
_{\sigma \pi }=0.15,\;\lambda _{\pi \sigma }=0.11,$ and $\lambda _{\pi \pi
}=0.21.$ These $\lambda $-values yield the mass-renormalization parameters
in Fig.~2 of Ref.~\onlinecite{Louie}: $m^{\ast }/m-1=\lambda _{\sigma
}=\lambda _{\sigma \pi }+\lambda _{\sigma \sigma }\approx 0.94$ and $\lambda
_{\pi }=\lambda _{\pi \sigma }+\lambda _{\pi \pi }\approx 0.32.$ The total
isotropic (thermodynamic) $\lambda =(N_{\sigma }\lambda _{\sigma }+N_{\pi
}\lambda _{\pi })/N=0.61$, which of course is the same as the one given by
Choi \emph{et al.} Here, and in Eq.~\ref{e1}, $N$ is the DOS summed over all
bands. With this two-gap model we have performed strong-coupling Eliashberg
calculations in order to compare the results for $T_{c}$ and the gaps with
those resulting from the fully anisotropic treatment. For all four spectral
functions we used the isotropic $\alpha ^{2}F(\omega )$ from 
Fig.~1 of Ref.~\onlinecite{Louie} scaled to produce the $\lambda $-matrix given above. 
The $\mu ^{\ast }(\omega _{c})$-matrix is obtained from Eq. \ref{e1} with $%
\lambda (\mathbf{k},\mathbf{k}^{\prime },0)$ substituted by Choi 
\emph{et al.}'s $\mu ^{\ast }(\omega _{c})$. The resulting $T_{c}$ and the gaps are
shown by dashed lines in Fig.~1 as functions of $\mu ^{\ast }(\omega _{c})$.
At $\mu ^{\ast }(\omega _{c})=0.12,$ as used by Choi \emph{et al.}, we get 
$T_{c}=43$ K, $\Delta _{\sigma }=7.2$ meV, and $\Delta _{\pi }=1.3$ meV. The
corresponding values quoted by Choi \emph{et al.} are 39 K, 6.8 meV and 1.8
meV. These differences are hardly due to intraband anisotropy, first of all
because it can only increase $T_{c}$. Secondly, increasing the number of
gaps from two to four in the Eliashberg equations, which should account for
most of the anisotropy beyond the two gap model, we found rather small
changes\cite{Note}.

If, on the other extreme, we assume that there is no Coulomb repulsion
between the $\sigma $ and $\pi $-electrons, then the corresponding two-gap
treatment gives the full lines in Fig.~1 and, hence, $T_{c}=38$ K, $\Delta
_{\sigma }=6.5$ meV, and $\Delta_{\pi }=1.8$ meV for $\mu ^{\ast }(\omega
_{c})=0.12$, incidentally, rather close to the values quoted 
in Refs.~\onlinecite{Louie,Choi}. If the magnitude of $\mu ^{\ast } $
in both calculations shown in Fig.~1 is adjusted to produce the 
same $T_c$ of 39 K, the value of the lower gap changes from $\approx $ 2 eV
(diagonal) to $\approx $ 0.4 eV (uniform).

That uniform and diagonal Coulomb pseudopotentials yield different results
is not surprising: The same total Eliashberg\ $\mu ^{\ast }$ in the uniform
case is distributed over intra- and interband terms so that the $\sigma
\sigma $-part of the pairing interaction suffers less than in the case of a
diagonal $\mu ^{\ast }$. $\lambda _{\sigma \sigma }$ is more important for
the critical temperature, and $\lambda _{\sigma \pi }$ for generating 
$\Delta _{\pi }$. For uniform $\mu ^{\ast }$, therefore, the $T_{c}$ and 
$\Delta _{\sigma }$ are larger, and $\Delta _{\pi }$ is much smaller.

\begin{figure}[tbp]
\epsfig{file=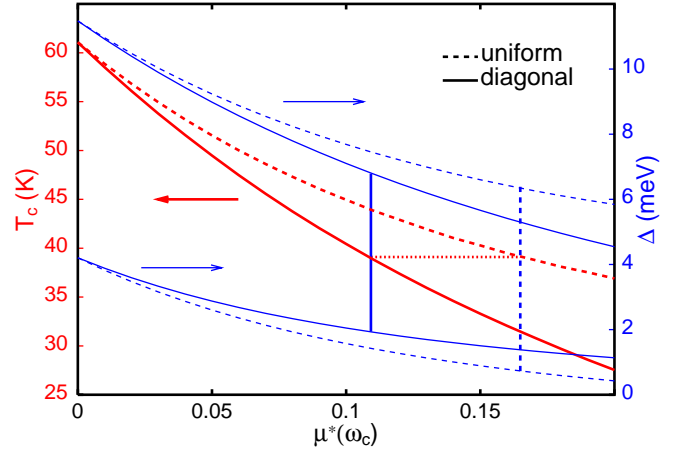,width=\linewidth,clip=true}
\caption{Critical temperature and the values of the $\protect\sigma $ and $%
\protect\pi $ gaps at 1 K as function of the renormalized Coulomb
pseudopotential, $\protect\mu ^{\ast }(\protect\omega _{c})$, in two models:
The uniform model where all matrix elements of the Coulomb repulsion are
equal and the diagonal model where the interband matrix elements are zero.
In both cases the normalization is chosen so as to produce given values of $%
\protect\mu ^{\ast }(\protect\omega _{c})$ after proper summation over all
bands. The two gaps obtained from 
$\protect\mu ^{\ast }(\protect\omega _{c})$'s giving $T_c$'s of 
39 K in the two models are connected vertically.}
\label{fig:tcvsmu}
\end{figure}

Having demonstrated that the assumed structure of $\mu ^{\ast }$ matters for
the details of the superconducting properties of MgB$_{2}$ \cite{Note2}, the
size of $\Delta _{\pi }$ in particular, let us finally estimate this
structure from first principles. The unrenormalized $\mu $ is the matrix
element $\left\langle n\mathbf{k\uparrow ,}n\,\mathbf{-k\downarrow }%
|V_{C}|n^{\prime }\mathbf{k}^{\prime }\uparrow ,n^{\prime }\,\mathbf{-k}%
^{\prime }\downarrow \right\rangle $ for scattering a Cooper pair from state 
$\left| n^{\prime }\mathbf{k}^{\prime }\right\rangle $ to state $\left| n%
\mathbf{k}\right\rangle $ via a phonon with wave-vector $\mathbf{k-k}%
^{\prime }.$ Inserting this matrix element in Eq. \ref{e1} instead of $%
\lambda (\mathbf{k},\mathbf{k}^{\prime }{,}0)$ yields $\mu _{nn^{\prime }}.$
Here $V_{C}(\mathbf{r,r}^{\prime })$ is the screened Coulomb interaction
between the electrons, and since it has short range in good metals, it makes
sense to take it proportional to the delta-function $\delta (\mathbf{r-r}%
^{\prime }).$ This leads to the following estimate:%
\begin{eqnarray}
\mu  &\varpropto &N\int \left| \psi _{n\mathbf{k}}(\mathbf{r})\right|
^{2}\left| \psi _{n^{\prime }\mathbf{k}^{\prime }}(\mathbf{r})\right|
^{2}d^{3}r  \nonumber \\
\mu _{nn^{\prime }} &\varpropto &N_{n^{\prime }}\int \left| \psi (\mathbf{r}%
)\right| _{n}^{2}\,\left| \psi (\mathbf{r})\right| _{n^{\prime
}}^{2}\,d^{3}r,  \label{e2}
\end{eqnarray}%
where $\left| \psi (\mathbf{r})\right| _{n}^{2}\equiv \sum_{\mathbf{k}%
}\left| \psi _{n\mathbf{k}}(\mathbf{r})\right| ^{2}\delta (\varepsilon _{n%
\mathbf{k}})/N_{n}$ is the shape, normalized to 1 in the cell or the
crystal, of the electron density of band $n$ at the Fermi level. These $%
\sigma $ and $\pi $ densities are shown in Fig.2, and they yield for the
ratios of the integrals in Eq.\ref{e2}%
\begin{equation}
\left\langle \left| \psi \right| _{\sigma }^{4}\right\rangle :\left\langle
\left| \psi \right| _{\pi }^{4}\right\rangle :\left\langle \left| \psi
\right| _{\sigma }^{2}\,\left| \psi \right| _{\pi }^{2}\right\rangle \sim
3.0:1.8:1.  \label{e3}
\end{equation}%
These ratios reflect the facts that the $\sigma $-density is more compact
than the $\pi $-density, and that the overlap of these two densities is
small. Note that the exceptional smallness of the interband impurity
scattering \cite{Mazin} in MgB$_{2}$ is due not only to this difference in
charge density, but also to a disparity of the $\sigma $ and $\pi $ \textit{%
wave functions}.

\begin{figure}[tbp]
\epsfig{file=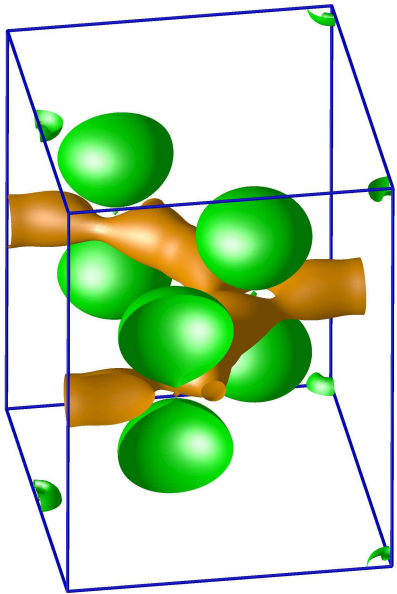,angle=0,width=.6\linewidth,clip=true}
\caption{Constant-density contour for the normalized $\protect\sigma $
(orange) and $\protect\pi $ (green) electron-densities, $\left| \protect\psi %
( \mathbf{r}) \right| _{\protect\sigma }^{2}$ and $\left| \protect\psi ( 
\mathbf{r}) \right| _{\protect\pi }^{2},$ at the Fermi level.}
\label{fig:2}
\end{figure}

From Eqs. \ref{e2}, \ref{e3} we get: 
$\mu_{\sigma\sigma }$:$\mu_{\pi\pi}$:$\mu_{\sigma\pi}$:$\mu_{\pi\sigma}$=3.1:2.6:1.4:1.
Now, any anisotropy in the \textit{bare} pseudopotential is further \emph{enhanced} in the
renormalized $\mu ^{\ast }$: In the one-band case $\mu $ is renormalized as 
$\mu ^{\ast }(\omega _{c})=\mu / \left[ 1+\mu \ln (W/\omega _{c})\right]$, 
where $W$ is a characteristic electronic energy of the order of the
bandwidth or plasma frequency. For the multi-band case, this is a matrix
equation with $W$ being a diagonal matrix with elements $W_{n}$. Assuming
for simplicity that $\mu _{\sigma \sigma }=\mu _{\pi \pi }=A \mu_{\sigma\pi }$ 
with $A>1$, and that $\mu _{\sigma \sigma
}\log (W_{\sigma }/\omega _{c})=\mu _{\pi \pi }\log (W_{\pi }/\omega _{c})=L,
$ one obtains: $A^{\ast }=A+(A-A^{-1})L$. For MgB$_{2},$ $L\sim 0.5-1$ and $%
A\sim 2.3,$ so that $A^{\ast }\sim 3-4,$ which is very different from the
uniform $\mu $. 

In conclusion: Any difference between the results of the fully anisotropic
Eliashberg formalism and those of the two-gap formalism will hardly be
observable in real MgB$_{2}$-samples. On the other hand, the anisotropy of
the Coulomb pseudopotential is likely to have an observable effect on the
size of the small gap, $\Delta _{\pi }$.

The authors thank W.E. Pickett for numerous helpful discussions, and for
critical reading of the manuscript.


\begin{thebibliography}{99}
\bibitem{gaps}  We have verified this analytical criterion numerically
using some simplified models for the electron-phonon coupling matrix.
\bibitem{Note} This is not surprising, given that the variation of the
calculated order parameter\cite{Choi}, as well as of the mass renormalization%
\cite{Louie}, within each band is at most 10\%.
\end{thebibliography}
\end{document}